\documentclass[12pt]{iopart}

\usepackage{iopams}

\usepackage{latexsym,epsfig,epstopdf}

\newcommand{\ket}[1]{\left | #1 \right\rangle}
\newcommand{\bra}[1]{\left \langle #1 \right |}

\newcommand{\proj}[1]{\ket{#1}\bra{#1}}

\newtheorem{lemma}{Lemma}
\newtheorem{theorem}{Theorem}

\begin{document}

\title[Werner gap in presence of simple coloured noise]{Werner gap in presence of simple coloured noise}

\author{Minh Cong Tran$^1$, Wies{\l}aw Laskowski$^2$ and Tomasz Paterek$^{1,3}$}

\address{$^1$ School of Physical and Mathematical Sciences, Nanyang Technological University, Singapore}
\address{$^2$ Institute of Theoretical Physics and Astrophysics, University of Gda\'nsk, Poland}
\address{$^3$ Centre for Quantum Technologies, National University of Singapore, Singapore}
\ead{cmtran@ntu.edu.sg}
\ead{wieslaw.laskowski@ug.edu.pl}
\ead{tomasz.paterek@ntu.edu.sg}

\begin{abstract}
A `Werner gap' is the range of relevant parameters characterising a quantum state for which it is both entangled and admits local hidden variable model.
Werner showed that the gap becomes maximal for entanglement mixed with white noise if subsystems have infinitely many levels.
Here we study pure entangled states mixed with simple coloured noise modelled as a single pure product state.
We provide an explicit local hidden variable model for quantum correlations of some states of this family and give hints that likely there is a model for all quantum predictions.
This demonstrates essentially maximal Werner gap already for two qubits.
Additionally to fundamental interest the study has implications for quantum computation and communication.
\end{abstract}


\maketitle

\section{Introduction}

Twenty five years separates two seminal contributions to foundations of physics being the topics of this special issue.
In 1964 Bell demonstrated that there is no local hidden variable (LHV) model (classical-like model) underlying statistical predictions of quantum mechanics~\cite{BELL1964}.
He proved this by constructing the (Bell) inequality that is satisfied by prediction of all the classical-like models and which is violated by quantum mechanical expectation values for suitable measurements performed on maximally entangled states.
It was clear that entanglement is a necessary ingredient for the violation, but the question whether its presence allows for the violation of some Bell inequality was left open for the twenty five years.
In 1989 Werner showed that in fact entanglement is not sufficient by constructing explicit LHV model for all quantum mechanical predictions of some entangled mixed states~\cite{WERNER}.

Both findings are clearly of great importance.
It is tempting to remove the statistical nature of quantum mechanics by introducing some underlying theory that keeps our classical mechanistic notions intact, whereas Bell's result proves this to be impossible.
On practical site, it gives tools to discriminate classical-like and quantum predictions and as such is used to reveal quantum superiority when solving certain tasks, e.g. cryptography~\cite{EKERT}, communication complexity~\cite{BUHRMAN} or computation~\cite{ANDERS}.
All these problems are related to computer science because predictions of LHV models can be seen as those arising from computations conducted on classical computers that are fed with data from a common source such as internet (but importantly the computers are not allowed to communicate with each other once the data is received).
Therefore, finding states violating Bell inequalities is also of practical interest and according to Werner's work for many tasks entanglement is not the source of quantum advantage.

Here we consider a class of mixed entangled states which nevertheless do not violate any Bell inequality for correlation functions.
The class studied contains a pure entangled state mixed with a pure product state.
All states of this family are known to be entangled even in the presence of infinitesimal admixture of infinitesimally small entanglement~\cite{HORODECKI}.
For completeness we provide an independent proof of this statement in~\ref{APP_ENT}.
We then focus on two qubits (spin-$\frac{1}{2}$ systems) and give a simple LHV model for maximally entangled state mixed with a specific product state that shows that no Bell inequality is violated for entanglement admixture at least up to $\frac{1}{2}$.
The model mimics all quantum correlations of the state as well as local expectation values for equatorial measurements on the Bloch sphere.
Next we extend the model to less entangled pure states.
Since smaller entanglement is being admixed one might intuitively expect that the range of its admixture compatible with the LHV model increases.
Indeed, we show this to be the case and in the limit of tiny entanglement, its admixture compatible with the LHV model approaches unity.
Therefore, we demonstrate that the Werner gap~\cite{LASKOWSKI}, i.e. the range of pure entanglement admixture for which the mixed entangled state admits LHV model, can be essentially maximal already for the system of two qubits.
For comparison, in his already mentioned seminal contribution Werner shows that his original model gives the maximal gap in the limit of infinitely-dimensional systems~\cite{WERNER}.
It should be emphasised that Werner's model mimics all quantum predictions whereas our model mimics only the correlations.
Nevertheless we conjecture that for the studied two-qubit states such a general model exists and provide hints why this should be the case.

Finally, we show a similarity between the cases of white and coloured noise.
In both cases known LHV models for two qubits cannot reach entanglement admixture for which known Bell inequalities are violated.
For coloured noise we show this by studying standard Bell inequalities for correlation functions~\cite{ZUK-BRUK, WW}.


\section{Werner gap}

Let us begin with the definition of the Werner gap and its brief and incomplete history.
The state introduced by Werner is invariant under the same unitary operation applied to both its $d$-level subsystems
and has the form~\cite{WERNER}:
\begin{equation}
\rho_{\mathrm{Werner}} = p \, \rho_{\mathrm{ent}} + (1-p) \, \frac{1}{d^2} \hat I.
\label{GEN_WERNER}
\end{equation}
The entangled state is an even mixture of all the singlet states $\ket{\psi_{kl}^-} = \frac{1}{\sqrt{2}}(\ket{kl} - \ket{lk})$ in two-dimensional subspaces
\begin{equation}
\rho_{\mathrm{ent}} = \frac{2}{d(d-1)} \sum_{k < l}^d \proj{\psi_{kl}^-},
\end{equation}
whereas the unentangled completely mixed state $\frac{1}{d^2} \hat I$ is called `white noise'.
The state (\ref{GEN_WERNER}) is entangled for
\begin{equation}
p > p_{\mathrm{ent}} = \frac{1}{d+1},
\end{equation}
and Werner constructed explicit LHV model for all projective measurements for admixtures
\begin{equation}
p \le p_{\mathrm{lhv}} = 1 - \frac{1}{d}.
\end{equation}
We define the `Werner gap' as the range of parameters describing the state for which it is both entangled and admits LHV model.
For the case of Eq.~(\ref{GEN_WERNER}) the Werner gap is therefore defined as
\begin{equation}
\Delta \equiv p_{\mathrm{lhv}} - p_{\mathrm{ent}}.
\end{equation}
In the simplest system of two qubits the gap is $\Delta = \frac{1}{2} - \frac{1}{3} = \frac{1}{6}$, and it reaches maximum achievable value $\Delta \to 1$ as $d \to \infty$.
As far as we know this is the only example of an entangled state admitting LHV model in the whole range of relevant parameters.
Almeida \emph{et al.}~\cite{ALMEIDA} found LHV model for the isotropic states which even for large $d$ shows positive Werner gap, but the gap is never greater than $\frac{1}{2}$ and it tends to zero for $d \to \infty$.
The results of Werner were improved by Acin \emph{et al.}~\cite{ACIN} who show a two-qubit LHV model for entanglement admixtures roughly below $p_{\mathrm{lhv}} < \frac{2}{3}$,
and using the results related to Grothendieck's constant show that for any mixed state, i.e. in the worst case scenario, the admixture of entanglement to white noise below which there is LHV model is in the limit of infinite dimension somewhere in between $0.5611$ and $0.5964$.
This latter result holds for quantum mechanical predictions for joint correlations of traceless two-outcome observables.
The Werner gap for correlations between pairs of dichotomic observables was studied in~\cite{LASKOWSKI},
and the present work is an extension of it to correlations (and some local expectation values) between arbitrary projective measurements, 
but yet it is not as general as those of Werner or Almeida \emph{et al}.


\section{States of interest}

We shall consider here mainly the states of two qubits of the following form:
\begin{equation}
\rho = p \proj{\psi_{\mathrm{ent}}} + (1-p) \proj{\psi_{\mathrm{prod}}},
\label{COLORED}
\end{equation}
where $\ket{\psi_{\mathrm{ent}}}$ is a pure entangled state, $\ket{\psi_{\mathrm{prod}}}$ is a pure product state, and $p$ is a probability characterising admixture of entanglement.
The Werner gap for this state is the range of $p$ for which both the state is entangled and admits LHV model for correlations between all possible projective measurements that can be performed upon it.
We first show that all such states are entangled as soon as $p > 0$, and next by specifying $\ket{\psi_{\mathrm{ent}}}$ and $\ket{\psi_{\mathrm{prod}}}$ we will present their LHV models
together with the range of settings for which it agrees with quantum predictions.


\section{Entanglement}

Any state of the form (\ref{COLORED}) is entangled for all $p>0$.
This has been effectively shown on several occasions by noticing that $\rho$ is of rank two.
Ref.~\cite{BADZIAG2002} proves that every entangled state of rank two has entangled $2 \times 2$ dimensional subspace (see also~\cite{HORODECKI}), 
and provides necessary and sufficient conditions for entanglement of such states in terms of decomposition of common eigenstate of concurrence matrices.
Another derivation of the if and only if conditions is presented in Ref.~\cite{FEI2003}.
Explicit expression for the amount of entanglement in such states is given in Ref.~\cite{OSBORNE2005} and in principle could be used for the proof.
Finally, a direct theorem characterising entanglement of states (\ref{COLORED}) is presented in Ref.~\cite{HORODECKI} and uses the criterion of positive partial transposition~\cite{PPT1,PPT2}.

For completeness, in ~\ref{APP_ENT} we give yet another proof of this statement for $N$ qubits that uses geometrical properties of the underlying Hilbert space.


\section{Local hidden variable model}

Despite being entangled some states (\ref{COLORED}) admit LHV model.
Let us first restrict our attention to the two-qubit states
\begin{equation}
\rho = p \proj{\psi^-} + (1-p) \proj{z+} \otimes \proj{z+},
\label{SINGLET}
\end{equation}
where $\ket{\psi^-} = \frac{1}{\sqrt{2}}(\ket{z+} \ket{z-} - \ket{z-} \ket{z+})$ is the Bell singlet state, and $\ket{z\pm}$ is the eigenstate of local $\sigma_z$ Pauli operator corresponding to the eigenvalue $\pm 1$.
For this state, quantum mechanics predicts that correlations between outcomes of local dichotomic measurements are given by:
\begin{eqnarray}
E_{\mathrm{QM}}(\vec a, \vec b) & = & - p \, \vec a \cdot \vec b + (1-p) \, a_z b_z \nonumber \\
& = & - p \, a_x b_x - p \, a_y b_y + (1-2p) \, a_z b_z,
\label{QM_CORR}
\end{eqnarray}
where $\vec a = (a_x,a_y,a_z)$ and $\vec b = (b_x,b_y,b_z)$ are the Bloch vectors parameterising the local observables.
For the local expectation values quantum predictions are:
\begin{eqnarray}
E_{\mathrm{QM}}(\vec a) & = & (1-p) \, a_z, \nonumber \\
E_{\mathrm{QM}}(\vec b) & = & (1-p) \, b_z.
\label{QM_LOC}
\end{eqnarray}
The LHV model for the quantum correlations of the states with $p \le \frac{1}{2}$ is very simple and we only need to realise that they are identical to those in the following separable state
\begin{eqnarray}
\rho_{\mathrm{lhv}} & = & \frac{p}{2} \proj{x+} \otimes \proj{x-} + \frac{p}{2} \proj{x-} \otimes \proj{x+} \nonumber \\
& + & \frac{p}{2} \proj{y+} \otimes \proj{y-} + \frac{p}{2} \proj{y-} \otimes \proj{y+} \nonumber \\
& + & (1- 2p) \proj{z+} \otimes \proj{z+} \quad \textrm{ for } \quad p \le 1/2.
\label{SEP_LHV}
\end{eqnarray}
Note also that the local expectation values agree with the quantum ones for equatorial measurements $a_z = b_z = 0$.
Although we were not able to construct explicit LHV model also covering the remaining local expectation values,
using the software of Ref.~\cite{GRUCA} we cannot find a violation of any Bell inequality with up to ten settings per side for $p < \frac{1}{\sqrt{2}}$.
While allowing more settings might lower a bit the value of the critical admixture it is very unlikely that it can be smaller that $\frac{1}{2}$.
Additional hint that there exists a model for all the quantum predictions of the states with $p \le \frac{1}{2}$ comes from the following theorem.
\begin{theorem}
If there exists LHV model for the state (\ref{SINGLET}) with $p=\frac{1}{2}$, then it can be used to construct LHV model for any $p \le \frac{1}{2}$. 
\end{theorem}
The feature that makes the state with $p = \frac{1}{2}$ special is the lack of correlations along local $z$ axes while having non-zero local expectation values
\begin{eqnarray}
E_{\frac{1}{2}}(\vec a, \vec b) & = & - \frac{1}{2} \, a_x b_x - \frac{1}{2} \, a_y b_y , \nonumber \\
E_{\frac{1}{2}}(\vec a) & = & \frac{1}{2} \, a_z, \nonumber \\
E_{\frac{1}{2}}(\vec b) & = & \frac{1}{2} \, b_z.
\end{eqnarray}
Consider this hypothetical protocol is used with probability $2q$, where $q < \frac{1}{2}$.
This gives
\begin{eqnarray}
E(\vec a, \vec b) & = & - q \, a_x b_x - q \, a_y b_y , \nonumber \\
E(\vec a) & = & q \, a_z, \nonumber \\
E(\vec b) & = & q \, b_z.
\end{eqnarray}
With the remaining $1-2q$ probability the source sends a product state $\ket{z+}\ket{z+}$
and locally the algorithm to calculate the outcomes is just to follow the usual quantum rules.
This exactly adds the missing terms
\begin{eqnarray}
E_q(\vec a, \vec b) & = & - q \, a_x b_x - q \, a_y b_y + (1-2q) \, a_z b_z, \nonumber \\
E_q(\vec a) & = & (1-q) \, a_z , \nonumber \\
E_q(\vec b) & = & (1-q) \, b_z. \qquad  \textrm{ } \qquad  \textrm{ } \qquad \square
\end{eqnarray}
Consider now the following two-parameter family of states
\begin{equation}
\rho = p \proj{\psi_\xi} + (1-p) \proj{z+} \otimes \proj{z+},
\label{GEN}
\end{equation}
where the entangled state is of the form
\begin{equation}
\ket{\psi_\xi} = \cos \xi \ket{z+} \ket{z-} - \sin \xi \ket{z-} \ket{z+}, \quad \textrm{ with } \quad \xi \in [0, \pi/4].
\label{GEN_PSI}
\end{equation}
We study its correlations only. 
Quantum mechanics predicts
\begin{eqnarray}
E_{\mathrm{QM}}(\vec a, \vec b) & = & - p \sin (2 \xi) \, a_x b_x - p \sin (2 \xi) \, a_y b_y + (1-2p) \, a_z b_z.
\end{eqnarray}
Using the same method as above, the separable state of the form (\ref{SEP_LHV}) with the weights $p \sin (2 \xi)$, $p \sin (2 \xi)$, and $1-2p$, provides the LHV model for
\begin{equation}
p_{\mathrm{lhv}} \le \frac{1}{1+\sin(2 \xi)}.
\end{equation}
Therefore, for very weakly entangled states, as $\xi \to 0$, even very high admixture of entanglement has correlations compatible with the LHV model, $p_{\mathrm{lhv}} \to 1$.
Since the state is entangled for all $p_{\mathrm{ent}} > 0$, the Werner gap essentially reaches its maximal value $\Delta \to 1$.

\begin{figure}
\begin{center}
\qquad \qquad \includegraphics[scale=0.4]{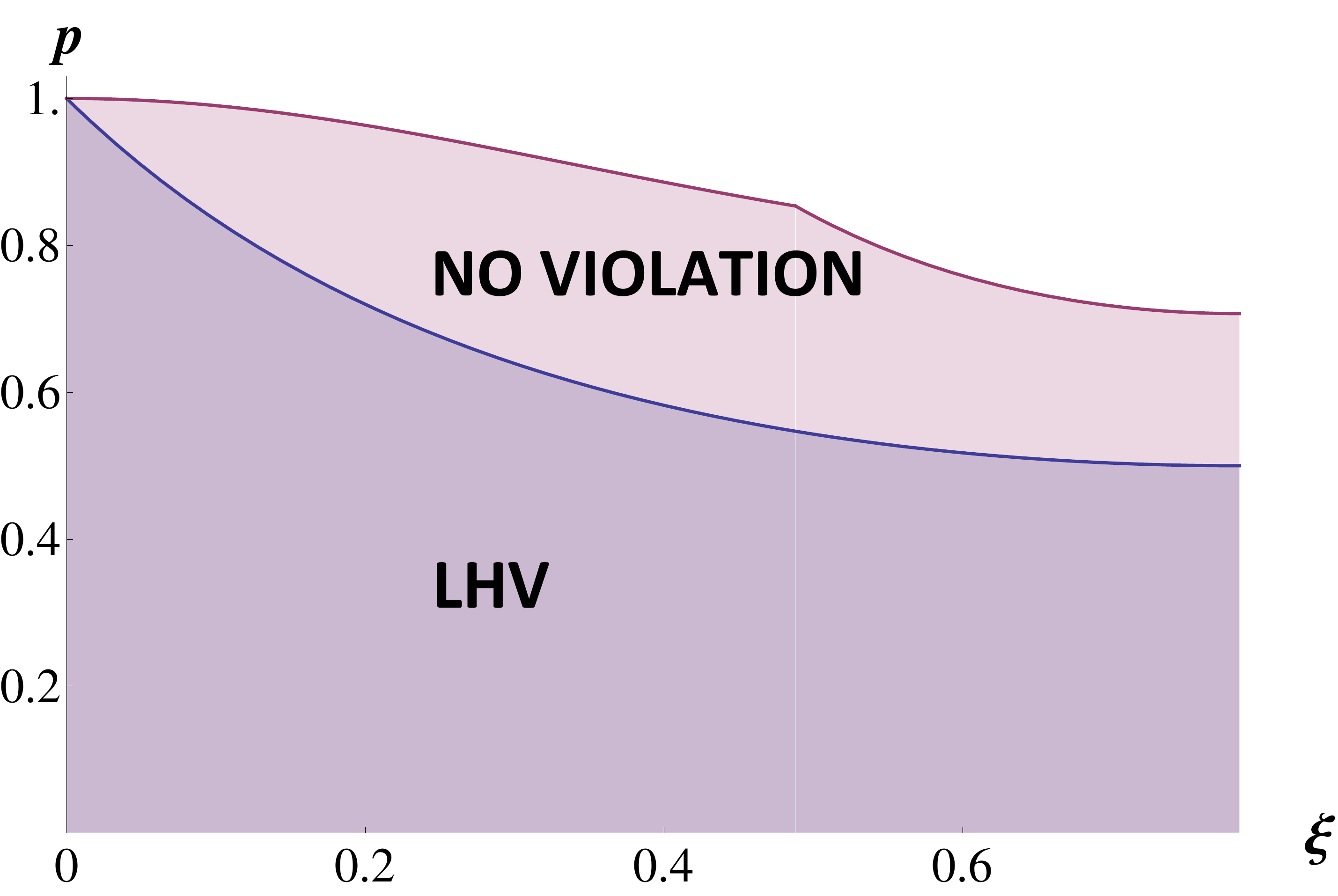}
\end{center}
\caption{Entangled and local states. In the main text we describe a local hidden variable model that mimicks the quantum mechanical correlations of state (\ref{GEN}) in the region marked as LHV.
The states of this family are entangled for all $p>0$ and $\xi > 0$.
Therefore, if $\xi$ is tiny the entangled states (\ref{GEN}) admit the LHV model essentially for the range of entanglement admixtures approaching $1$ (maximal Werner gap).
We also mark in the figure the region in which all standard (two-setting) Bell inequalities for correlation functions are not violated.
As for the case of admixed white noise, also here there is still discrepancy between these two sets.}
 \label{FIG1}
\end{figure}
Finally, we show using standard Bell inequalities for correlation functions~\cite{ZUK-BRUK,WW} that despite essentially maximal Werner gap the LHV model does not recover all the range where the inequalities are not violated.
Therefore, either inequalities with higher number of settings improve on the set of states allowing the violation or there exists LHV model for the bigger set of states.
The necessary and sufficient condition for violation of a complete set of such inequalities is given by~\cite{ZUK-BRUK}:
\begin{equation}
E^2(\vec a_1, \vec b_1) + E^2(\vec a_2, \vec b_2) > 1,
\end{equation}
where the sum on the left-hand side involves two biggest squared correlations in a state, measured along bi-orthogonal axes.
For our state, either $E^2(\vec a_1, \vec b_1) = p^2 \sin^2(2 \xi)$ and $E^2(\vec a_2, \vec b_2) = (1-2p)^2$, or $E^2(\vec a_1, \vec b_1) = E^2(\vec a_2, \vec b_2) = p^2 \sin^2(2 \xi)$,
hence the state violates some two-setting correlation Bell inequality for
\begin{eqnarray}
p & > & \min \left(\frac{1}{\sqrt{2} \sin(2 \xi)}, \frac{4}{4 + \sin^2(2 \xi)} \right) \\
& = & 
\Bigg\{
\begin{array}{rcl}
\frac{1}{\sqrt{2} \sin(2 \xi)} & \textrm{ for } & \xi \in [\xi_1,\xi_2], \\
\frac{4}{4 + \sin^2(2 \xi)}) & \textrm{ otherwise }
\end{array}
\end{eqnarray}
where $\xi_1 \approx 0.16 \pi$ and $\xi_2 \approx 0.34 \pi$.
The range of parameters $\xi$ and $p$ where the states do not admit the LHV model for correlations and yet do not violate the standard correlation Bell inequalities is depicted in Fig.~\ref{FIG1}.


\section{Conclusions}

We presented a simple state for which the Werner gap, i.e. the range of relevant parameters for which the state is both entangled and allows for local hidden variable model, is maximal.
Our example involves two qubits whereas the only other case of maximal Werner gap known to us involves two systems of dimensionality approaching infinity.
The model presented mimics quantum correlations between all projective measurements and local expectation values for equatorial measurements,
and it is argued that likely a local hidden variable model exists for all quantum predictions.
All this shows that ability to violate a Bell inequality is highly dependent on the type of noise in the experiment~\cite{ LASKOWSKI,CABELLO, SEEVINCK, AOLITA}
and therefore tasks such as communication complexity, with performance directly linked to the amount of Bell violation~\cite{BUHRMAN},
should be executed in well controlled environments.

\ack

This work is supported by the National Research Foundation, Ministry of Education of Singapore, start-up grant of the Nanyang Technological University, and NCN Grant No. 2012/05/E/ST2/02352.

\appendix


\section{Pure entanglement mixed with pure product state}
\label{APP_ENT}
\setcounter{section}{1}

We prove here for $N$ qubits that tiny admixture of even infinitesimal entanglement to a product state always gives a mixed entangled state.
Our main theorem uses the lemma being Theorem $1$ of Ref.~\cite{SANPERA1998} stating that for any plane in $\mathbb{C}^d \otimes \mathbb{C}^d$ defined by two product vectors, either all the states in this plane are product vectors, or there is no other product vector in it.
For completeness, we also prove this lemma.

\begin{lemma}
\label{LEM}
Let $\ket{\Psi}$ be arbitrary pure product state of $N$ qubits. Then the superposition
\begin{equation}
	\ket{\Psi'}=\alpha\ket{00\dots 0}+\beta\ket{\Psi}, \label{sup}
\end{equation}
is either a product state for any complex $\alpha, \beta$, or an entangled state for any complex $\alpha, \beta\neq 0,1$.
\end{lemma}
The cases of $\alpha=0,1$ are trivial.
Consider $\alpha\neq 0,1$, and hence $\beta\neq 0,1$ as well. 
By assumption $\ket{\Psi}$ is a product state:
\begin{equation}
	\ket{\Psi}=a\left(\ket{0}+k_1\ket{1}\right)\otimes \left(\ket{0}+k_2\ket{1}\right)\otimes\dots\otimes \left(\ket{0}+k_N\ket{1}\right),
\end{equation}
where $a$ is a normalization constant.
Clearly, if only one $k_j$ is non-zero the superposed state $\ket{\Psi'}$ is a product state for all $\alpha$ and $\beta$.
We show that if at least two coefficients $k_j$ are non-zero, say $k_1$ and $k_2$, the superposed state $\ket{\Psi'}$ is entangled for all $\alpha, \beta\neq 0,1$.
Assume by contradiction that there exists $\alpha$ and $\beta$ such that $\ket{\Psi'}$ is a product state. Then 
\begin{eqnarray}
	\ket{\Psi'}&=a'\left(\ket{0}+k'_1\ket{1}\right)\otimes \left(\ket{0}+k'_2\ket{1}\right)\otimes\dots\otimes \left(\ket{0}+k'_N\ket{1}\right)\\
	&=\alpha\ket{00...0}+\beta a \left(\ket{0}+k_1\ket{1}\right)\otimes \left(\ket{0}+k_2\ket{1}\right)\otimes\dots\otimes \left(\ket{0}+k_N\ket{1}\right), \nonumber
\end{eqnarray}
and the coefficients satisfy
\begin{eqnarray}	
\alpha+\beta a & = & a', \label{coeff1} \\
\beta a k_1 & = & a' k'_1, \label{coeff2} \\
\beta a k_2 & = & a' k'_2, \label{coeff3} \\
& \vdots & \nonumber \\
\beta a k_N&=a' k'_N, \label{coeffN} \\
\beta a k_1 k_2&=a' k'_1 k'_2, \label{coeff12} \\
& \vdots & \nonumber
\end{eqnarray}
Since $\beta, a, k_1$ are all nonzero, $k'_1$ is also nonzero. From (\ref{coeff2}) and (\ref{coeff12}) it follows that $k_2=k'_2$, 
from (\ref{coeff3}) one has $\beta a=a'$, and from (\ref{coeff1}) one has $\alpha=0$, which contradicts the assumption that $\alpha\neq 0$. 
$\square$

\begin{theorem}
The $N$-qubit state
\begin{equation}
\rho = \epsilon \proj{\psi_{\mathrm{ent}}} + (1-\epsilon) \proj{\psi_{\mathrm{prod}}}, 
\label{theoeqn2}
\end{equation}
where $\ket{\psi_{\mathrm{ent}}}$ and $\ket{\psi_{\mathrm{prod}}}$ are respectively entangled and product pure states, is entangled if and only if $\epsilon > 0$.
\end{theorem}
Without loss of generality we put $\ket{\psi_{\mathrm{prod}}} = \ket{00...0} \equiv \ket{\mathbf{0}}$, i.e. these product states define the computational basis, and set $\epsilon > 0$.
The pure entangled state can always be decomposed as 
\begin{eqnarray}
	\ket{\psi_{\mathrm{ent}}} = a \ket{\mathbf{0}} + b \ket{\mathbf{0}^{\perp}},
\end{eqnarray}
where $\ket{\mathbf{0}^{\perp}}$ is a state orthogonal to $\ket{\mathbf{0}}$ and $a, b$ are normalized complex coefficients. 
The density matrix $\rho$ in the two-dimensional subspace spanned by $\ket{\mathbf{0}}$ and $\ket{\mathbf{0}^{\perp}}$ reads:
\begin{eqnarray}
\rho=\left(
\begin{array}{cc}
(1-\epsilon)+\epsilon|a|^2 & \epsilon a b^* \\
\epsilon a^*b & \epsilon |b|^2
\end{array}
\right).
\label{rho1}
\end{eqnarray}
We shall proceed by contradiction. Assume that $\rho$ is separable:
\begin{eqnarray}
	\rho=\sum_{i=1} \lambda_i \ket{\Psi_i}\bra{\Psi_i},
\end{eqnarray}
where $\lambda_i$ is the probability of pure product state $\ket{\Psi_i}$.
For non-trivial separable state there must be at least two linearly independent product states and they can be used to decompose any other state in the two-dimensional subspace $\mathcal{S}$, spanned by $\ket{\mathbf{0}}$ and $\ket{\mathbf{0}^{\perp}}$, being the support of $\rho$.
Since the subspace $\mathcal{S}$ contains the entangled state $\ket{\psi_{\mathrm{ent}}}$, by Lemma~\ref{LEM} there are only two product states in $\mathcal{S}$.
One of them is $\ket{\Psi_1} = \ket{\mathbf{0}}$, and the other one must be of the form $\ket{\Psi_2} = \alpha \ket{\mathbf{0}} + \beta \ket{\mathbf{0}^\perp}$. Hence,
\begin{eqnarray}
	\rho & = & (1-\lambda)\ket{\Psi_1}\bra{\Psi_1}+\lambda\ket{\Psi_2}\bra{\Psi_2},\\
	& = & \left(
	\begin{array}{cc}
	1-\lambda+\lambda\alpha\alpha^* & \lambda\alpha\beta^* \\
	\lambda\beta\alpha^* & \lambda\beta\beta^*
	\end{array}
	\right).
	\label{rho2}
\end{eqnarray}
Comparing (\ref{rho1}) and (\ref{rho2}), one finds
\begin{eqnarray}
	\frac{b}{a}=\frac{\beta}{\alpha}.
\end{eqnarray}
Hence $\ket{\psi_{\mathrm{ent}}}$ is the same state as $\ket{\Psi_2}$ contradicting the assumption that $\ket{\psi_{\mathrm{ent}}}$ is entangled,
and we conclude that $\rho$ cannot be separable.
$\square$

\end{document}